\magnification \magstep1
\raggedbottom
\openup 2\jot 
\voffset6truemm
\def\cstok#1{\leavevmode\thinspace\hbox{\vrule\vtop{\vbox{\hrule\kern1pt
\hbox{\vphantom{\tt/}\thinspace{\tt#1}\thinspace}}
\kern1pt\hrule}\vrule}\thinspace}
\centerline {\bf A THEORY OF QUANTUM GRAVITY}
\centerline {\bf FROM FIRST PRINCIPLES}
\vskip 1cm
\noindent
Giampiero Esposito
\vskip 0.3cm
\noindent
{\it Istituto Nazionale di Fisica Nucleare, Sezione di Napoli,
Complesso Universitario di Monte S. Angelo, Via Cintia, Edificio N',
80126 Napoli, Italy}
\vskip 0.3cm
\noindent
{\it Dipartimento di Scienze Fisiche, Complesso Universitario di
Monte S. Angelo, Via Cintia, Edificio N', 80126 Napoli, Italy}
\vskip 1cm
\noindent
{\bf Abstract.} When quantum fields are studied on manifolds with
boundary, the corresponding one-loop quantum theory for bosonic
gauge fields with linear covariant gauges needs the assignment of
suitable boundary conditions for elliptic differential operators of
Laplace type. There are however deep reasons to modify such a scheme
and allow for pseudo-differential boundary-value 
problems. When the boundary operator is
allowed to be pseudo-differential while remaining a projector, the
conditions on its kernel leading to strong ellipticity of the
boundary-value problem are studied in detail. This makes it possible
to develop a theory of one-loop quantum gravity from first principles
only, i.e. the physical principle of invariance under infinitesimal
diffeomorphisms and the mathematical requirement of a strongly
elliptic theory.
\vskip 100cm
The space-time approach to quantum mechanics and quantum field theory
has led to several deep developments in the understanding of quantum
theory and space-time structure at very high energies.$^{1,2}$ In 
particular, we are here concerned with the choice of boundary conditions.
On using path integrals, which lead, in principle, to the appropriate
formulation of the ideas of Feynman, DeWitt 
and many other authors,$^{2-5}$
the assignment of boundary conditions consists of two main steps:
\vskip 0.3cm
\noindent
(i) Choice of Riemannian geometries and field configurations to be
included in the path-integral representation of transition amplitudes.
\vskip 0.3cm
\noindent
(ii) Choice of boundary data to be imposed on the hypersurfaces
$\Sigma_{1}$ and $\Sigma_{2}$ bounding the given space-time region.
\vskip 0.3cm
\noindent
The main object of our investigation is the second problem of such
a list, when a one-loop approximation is studied for a bosonic
gauge theory in linear covariant gauges. The well posed mathematical
formulation relies on the ``Euclidean approach'', i.e., in geometric
language, on the use of differentiable manifolds endowed with 
positive-definite metrics $g$, so that space-time is actually
replaced by an $m$-dimensional Riemannian space $(M,g)$. 

In particular,
in Euclidean quantum gravity, mixed boundary conditions on metric
perturbations $h_{cd}$ occur naturally if one requires their
complete invariance under infinitesimal diffeomorphisms, as is
proved in detail in Ref. 6. On denoting by $N^{a}$ the inward-pointing
unit normal to the boundary, by
$$
q_{\; b}^{a} \equiv \delta_{\; b}^{a}-N^{a}N_{b}
\eqno (1)
$$
the projector of tensor fields 
onto $\partial M$, with associated projection operator
$$
\Pi_{ab}^{\; \; \; cd} \equiv q_{\; (a}^{c} \; q_{\; b)}^{d},
\eqno (2)
$$
the gauge-invariant boundary conditions for one-loop quantum gravity
read$^{6}$
$$
\Bigr[\Pi_{ab}^{\; \; \; cd}h_{cd}\Bigr]_{\partial M}=0,
\eqno (3)
$$
$$
\Bigr[\Phi_{a}(h)\Bigr]_{\partial M}=0,
\eqno (4)
$$
where $\Phi_{a}$ is the gauge-averaging functional necessary to
obtain an invertible operator $P_{ab}^{\; \; \; cd}$ on metric
perturbations. When $P_{ab}^{\; \; \; cd}$ is chosen to be of
Laplace type, $\Phi_{a}$ reduces to the familiar de Donder term
$$
\Phi_{a}(h)=\nabla^{b}\Bigr(h_{ab}-{1\over 2}g_{ab}g^{cd}
h_{cd}\Bigr)=E_{a}^{\; bcd}\nabla_{b}h_{cd},
\eqno (5)
$$
where $E^{abcd}$ is the DeWitt supermetric on the vector bundle
of symmetric rank-two tensor fields over $M$
($g$ being the metric on $M$):
$$
E^{abcd} \equiv {1\over 2}\Bigr(g^{ac}g^{bd}+g^{ad}g^{bc}
-g^{ab}g^{cd}\Bigr).
\eqno (6)
$$
The boundary conditions (3) and (4) can then be cast in the
Grubb--Gilkey--Smith form:$^{7,8}$
$$
\pmatrix{\Pi & 0 \cr \Lambda & I-\Pi \cr}
\pmatrix{[\varphi]_{\partial M} \cr [\varphi_{;N}]_{\partial M}}=0.
\eqno (7)
$$
However, the
work in Ref. 6 has shown that an operator of Laplace type on
metric perturbations is then incompatible with the requirement 
of strong ellipticity of the boundary-value problem, 
because the operator $\Lambda$ contains tangential derivatives
of metric perturbations.

To take care of this serious drawback, the work in Ref. 9 has 
proposed to consider in the boundary condition (4) a 
gauge-averaging functional given by the de Donder term (5) plus
an integro-differential operator on metric perturbations, i.e.
$$
\Phi_{a}(h) \equiv E_{a}^{\; bcd}\nabla_{b}h_{cd}
+\int_{M}\zeta_{a}^{\; cd}(x,x')h_{cd}(x')dV'.
\eqno (8)
$$
We now begin by remarking that the resulting boundary conditions 
can be cast in the form
$$
\pmatrix{\Pi & 0 \cr \Lambda+{\widetilde \Lambda} & I-\Pi \cr}
\pmatrix{[\varphi]_{\partial M} \cr [\varphi_{;N}]_{\partial M} \cr}
=0,
\eqno (9)
$$
where $\widetilde \Lambda$ reflects the occurrence of the integral
over $M$ in Eq. (8). It is convenient to work first in a
general way and then consider the form taken by these operators 
in the gravitational case. On requiring that the resulting 
boundary operator
$$
{\cal B} \equiv \pmatrix{\Pi & 0 \cr \Lambda+{\widetilde \Lambda}
& I-\Pi \cr}
\eqno (10)
$$
should remain a projector: ${\cal B}^{2}={\cal B}$, we find
the condition
$$
(\Lambda+{\widetilde \Lambda})\Pi
-\Pi(\Lambda+{\widetilde \Lambda})=0,
\eqno (11)
$$
which reduces to
$$
\Pi {\widetilde \Lambda}={\widetilde \Lambda}\Pi,
\eqno (12)
$$
by virtue of the property $\Pi \Lambda=\Lambda \Pi=0$ considered
in Ref. 6.

In Euclidean quantum gravity at one-loop level, Eq. (12) 
leads to
$$ 
\Pi_{a \; \; c}^{\; b \; \; r}(x) \int_{M}
\zeta_{b}^{\; cq}(x,x')h_{qr}(x')dV' 
=\int_{M}\zeta_{a}^{\; cd}(x,x')\Pi_{cd}^{\; \; \; qr}(x')
h_{qr}(x')dV',
\eqno (13)
$$   
which can be re-expressed in the form
$$
\int_{M}\left[\Pi_{a \; \; c}^{\; b \; \; r}(x)
\zeta_{b}^{\; cq}(x,x')-\zeta_{a}^{\; cd}(x,x')
\Pi_{cd}^{\; \; \; qr}(x')\right]h_{qr}(x')dV'=0.
\eqno (14)
$$
Since this should hold for all $h_{qr}(x')$, it eventually leads
to the vanishing of the term in square brackets in the integrand.
The notation $\zeta_{b}^{\; cq}(x,x')$ is indeed rather
awkward, because there is an even number of arguments, i.e.
$x$ and $x'$, with an odd number of indices. Hereafter, we
therefore assume that a vector field $T$ and kernel
$\widetilde \zeta$ exist such that
$$
\zeta_{b}^{\; cq}(x,x') \equiv T^{p}(x)
{\widetilde \zeta}_{bp}^{\; \; \; cq}(x,x')
\equiv T^{p}{\widetilde \zeta}_{bp}^{\; \; \; c'q'}.
\eqno (15)
$$
The projector condition (12) is therefore satisfied if and 
only if$^{10}$
$$
T^{p}(x)\left[\Pi_{a \; \; c}^{\; b \; \; r}(x)
{\widetilde \zeta}_{bp}^{\; \; \; cq}(x,x')
-{\widetilde \zeta}_{ap}^{\; \; \; cd}(x,x')
\Pi_{cd}^{\; \; \; qr}(x')\right]=0.
\eqno (16)
$$

We are now concerned with the issue of ellipticity of the
boundary-value problem of one-loop quantum 
gravity. For this purpose, we begin
by recalling what is known about ellipticity of the Laplacian 
(hereafter $P$) on a Riemannian manifold with smooth boundary.
This concept is studied in terms of the leading symbol of $P$.
It is indeed well known that the Fourier transform makes it
possible to associate to a differential operator of order $k$
a polynomial of degree $k$, called the characteristic polynomial
or symbol. The leading symbol, $\sigma_{L}$, picks out the
highest order part of this polynomial. For the Laplacian,
it reads
$$
\sigma_{L}(P;x,\xi)=|\xi|^{2}I=g^{\mu \nu}\xi_{\mu}\xi_{\nu}I.
\eqno (17)
$$
With a standard notation, $(x,\xi)$ are local coordinates
for $T^{*}(M)$, the cotangent bundle of $M$. The leading symbol
of $P$ is trivially elliptic in the interior of $M$, since the
right-hand side of (17) is positive-definite, and one has
$$
{\rm det}\Bigr[\sigma_{L}(P;x,\xi)-\lambda \Bigr]
=(|\xi|^{2}-\lambda)^{{\rm dim} \; V} \not = 0,
\eqno (18)
$$
for all $\lambda \in {\cal C}-{\bf R}_{+}$. In the presence of
a boundary, however, one needs a more careful definition of
ellipticity. First, for a manifold $M$ of dimension $m$, the
$m$ coordinates $x$ are split into $m-1$ local coordinates on
$\partial M$, hereafter denoted by $\left \{ {\hat x}^{k} 
\right \}$, and $r$, the geodesic distance to the boundary. 
Moreover, the $m$ coordinates $\xi_{\mu}$ are split into
$m-1$ coordinates $\left \{ \zeta_{j} \right \}$ (with $\zeta$
being a cotangent vector on the boundary), jointly with a real
parameter $\omega \in T^{*}({\bf R})$. At a deeper level, all this
reflects the split
$$
T^{*}(M)=T^{*}({\partial M})\oplus T^{*}({\bf R})
\eqno (19)
$$
in a neighbourhood of the boundary.$^{6,11}$

The ellipticity we are interested in requires now that $\sigma_{L}$
should be elliptic in the interior of $M$, as specified before, and
that strong ellipticity should hold. This means that a unique
solution exists of the differential equation obtained from
the leading symbol:
$$
\left[\sigma_{L}\left(P; \left \{ {\hat x}^{k} \right \}, r=0,
\left \{ \zeta_{j} \right \}, \omega \rightarrow -i
{\partial \over \partial r} \right)-\lambda \right]
\varphi(r,{\hat x},\zeta;\lambda)=0,
\eqno (20)
$$
subject to the boundary conditions
$$
\sigma_{g}(B)\left( \left \{ {\hat x}^{k} \right \},
\left \{ \zeta_{j} \right \} \right) \psi(\varphi)
=\psi'(\varphi)
\eqno (21)
$$
and to the asymptotic condition
$$
\lim_{r \to \infty}\varphi(r,{\hat x},\zeta;\lambda)=0.
\eqno (22)
$$
In Eq. (21), $\sigma_{g}$ is the {\it graded leading symbol} of the
boundary operator in the local coordinates
$\left \{ {\hat x}^{k} \right \}, \left \{ \zeta_{j} \right \}$, 
and is given by
$$
\sigma_{g}(B)=\pmatrix{\Pi & 0 \cr i \Gamma^{j}\zeta_{j}
& I-\Pi \cr}.
\eqno (23)
$$
Roughly speaking, the above
construction uses Fourier transform and the
inward geodesic flow to obtain the ordinary differential
equation (20) from the Laplacian, with corresponding Fourier
transform (21) of the original boundary conditions.  
The asymptotic condition (22) picks out the solutions of Eq. (20)
which satisfy Eq. (21) with arbitrary boundary 
data $\psi'(\varphi) \in C^{\infty}(W',{\partial M})$ 
for $W'$ a vector bundle over the boundary, and
vanish at infinite geodesic distance to the boundary. When all
the above conditions are satisfied $\forall \zeta \in 
T^{*}({\partial M}), \forall \lambda \in {\cal C}-{\bf R}_{+},
\forall (\zeta,\lambda) \not = (0,0)$ and $\forall \psi'(\varphi)
\in C^{\infty}(W',{\partial M})$, the boundary-value problem
$(P,B)$ for the Laplacian is said to be strongly elliptic with
respect to the cone ${\cal C}-{\bf R}_{+}$. 

However, when the gauge-averaging functional (8) is used in the
boundary condition (5), the work in Ref. 9 has proved that the
operator on metric perturbations takes the form of an operator of
Laplace type $P_{ab}^{\; \; \; cd}$ plus an integral operator
$G_{ab}^{\; \; \; cd}$. Explicitly, one finds$^{9}$ (with
$R_{\; bcd}^{a}$ being the Riemann curvature of the background 
geometry $(M,g)$)
$$ 
P_{ab}^{\; \; \; cd}=E_{ab}^{\; \; \; cd}(-\cstok{\ }+R)
-2 E_{ab}^{\; \; \; qf}R_{\; qpf}^{c} g^{dp}
-E_{ab}^{\; \; \; pd} R_{p}^{\; c} 
-E_{ab}^{\; \; \; cp}R_{p}^{\; d},
\eqno (24) 
$$
$$
G_{ab}^{\; \; \; cd}=U_{ab}^{\; \; \; cd}
+V_{ab}^{\; \; \; cd},
\eqno (25)
$$
where
$$
U_{ab}^{\; \; \; cd}h_{cd}(x)=-2E_{rsab}\nabla^{r} \int_{M}
T^{p}(x){\widetilde \zeta}_{\; p}^{s \; \; cd}(x,x')h_{cd}(x')dV',
\eqno (26)
$$
$$
h^{ab}V_{ab}^{\; \; \; cd}h_{cd}(x)=\int_{M^{2}}h^{ab}(x')
T^{q}(x){\widetilde \zeta}_{pqab}(x,x')T^{r}(x)
{\widetilde \zeta}_{\; r}^{p \; \; cd}(x,x'')h_{cd}(x'')dV'dV''.
\eqno (27)
$$

We now assume that the operator on metric perturbations, which is
so far an integro-differential operator defined by a kernel, is 
also pseudo-differential. This means that it can be characterized 
by suitable regularity properties obeyed by the symbol. More 
precisely, let $S^{d}$ be the set of all symbols $p(x,\xi)$ such
that 
\vskip 0.3cm
\noindent
(1) $p$ is $C^{\infty}$ in $(x,\xi)$, with compact $x$ support.
\vskip 0.3cm
\noindent
(2) For all $(\alpha,\beta)$, there exist constants $C_{\alpha,\beta}$
for which
$$ \eqalignno{
\; & \left | (-i)^{\sum_{k=1}^{m}(\alpha_{k}+\beta_{k})}
{\left({\partial \over \partial x_{1}}\right)^{\alpha_{1}}
... \left({\partial \over \partial x_{m}}\right)}^{\alpha_{m}}
{\left({\partial \over \partial \xi_{1}}\right)^{\beta_{1}}
... \left({\partial \over \partial \xi_{m}}\right)}^{\beta_{m}}
p(x,\xi)\right | \cr
& \leq C_{\alpha,\beta}
{\left(1+\sqrt{g^{ab}(x)\xi_{a}\xi_{b}}\right)}^{d-\sum_{k=1}^{m}\beta_{k}},
&(28)\cr}
$$
for some {\it real} (not necessarily positive) value of $d$. The
associated pseudo-differential operator, defined on the Schwarz space
and taking values in the set of smooth functions on $M$ with compact
support:
$$
P: {\cal S} \rightarrow C_{c}^{\infty}(M)
$$
acts according to
$$
Pf(x) \equiv \int e^{i(x-y)\cdot \xi}p(x,\xi)f(y)\mu(y,\xi),
\eqno (29)
$$
where $\mu(y,\xi)$ is here meant to be the invariant integration
measure with respect to $y_{1},...,y_{m}$ and
$\xi_{1},...,\xi_{m}$. Actually, one first gives the definition 
for pseudo-differential operators $P: {\cal S} \rightarrow
C_{c}^{\infty}({\bf R}^{m})$, eventually proving that a
coordinate-free definition can be given and extended to smooth
Riemannian manifolds.$^{11}$

In the presence of pseudo-differential operators, both ellipticity
in the interior of $M$ and strong ellipticity of the 
boundary-value problem need a more involved formulation. In our
paper, inspired by the flat-space analysis in Ref. 12, we make
the following requirements.$^{10}$
\vskip 5cm
\leftline {\bf (i) Ellipticity in the Interior}
\vskip 0.3cm
\noindent
Let $U$ be an open subset with compact closure in $M$, and
consider an open subset $U_{1}$ whose closure ${\overline U}_{1}$
is properly included into $U$: ${\overline U}_{1} \subset U$.
If $p$ is a symbol of order $d$ on $U$, it is said to be
{\it elliptic} on $U_{1}$ if there exists an open set $U_{2}$
which contains ${\overline U}_{1}$ and positive constants
$C_{0},C_{1}$ so that
$$
|p(x,\xi)|^{-1} \leq C_{1} (1+|\xi|)^{-d},
\eqno (30)
$$
for $|\xi| \geq C_{0}$ and $x \in U_{2}$, where $|\xi| \equiv
\sqrt{g^{ab}(x)\xi_{a}\xi_{b}}$. The corresponding operator $P$
is then elliptic.
\vskip 0.3cm
\leftline {\bf (ii) Strong Ellipticity in the Absence of Boundaries}
\vskip 0.3cm
\noindent
Let us assume that the symbol under consideration is
{\it polyhomogeneous}, in that it admits an asymptotic expansion
of the form
$$
p(x,\xi) \sim \sum_{l=0}^{\infty}p_{d-l}(x,\xi),
\eqno (31)
$$
where each term $p_{d-l}$ has the {\it homogeneity property}
$$
p_{d-l}(x,t\xi)=t^{d-l}p_{d-l}(x,\xi) \; \; {\rm if} \; \;
t \geq 1 \; \; {\rm and} \; \; |\xi| \geq 1.
\eqno (32)
$$
The leading symbol is then, by definition,
$$
p^{0}(x,\xi) \equiv p_{d}(x,\xi).
\eqno (33)
$$
Strong ellipticity in the absence of boundaries is formulated in
terms of the leading symbol, and it requires that
$$
{\rm Re} \; p^{0}(x,\xi) \geq c(x) |\xi|^{d},
\eqno (34)
$$
where $x \in M$ and $|\xi| \geq 1$, $c$ being a positive function
on $M$. It can then be proved that the G\"{a}rding inequality holds,
according to which, for any $\varepsilon >0$,
$$
{\rm Re}(Pu,u) \geq b { \left \| u \right \| }_{{d\over 2}}^{2}
-b_{1}{ \left \| u \right \| }_{{d\over 2}-\varepsilon}^{2}
\; \; {\rm for} \; \; u \in H^{{d\over 2}}(M),
\eqno (35)
$$
with $b>0$.
\vskip 0.3cm
\leftline {\bf (iii) Strong Ellipticity in the Presence of Boundaries}
\vskip 0.3cm
\noindent
The homogeneity property (32) only holds for $t \geq 1$ and
$|\xi| \geq 1$. Consider now the case $l=0$, for which one obtains
the leading symbol which plays the key role in the definition 
of ellipticity. If $p^{0}(x,\xi) \equiv p_{d}(x,\xi) \equiv
\sigma_{L}(P;x,\xi)$ is not a polynomial (which corresponds
to the genuinely pseudo-differential case) while being a homogeneous
function of $\xi$, it is irregular at $\xi=0$. When $|\xi| \leq 1$,
the only control over the leading symbol is provided by
estimates of the form$^{12}$
$$ \eqalignno{
\; & \left | (-i)^{\sum_{k=1}^{m}(\alpha_{k}+\beta_{k})}
{\left({\partial \over \partial x_{1}}\right)}^{\alpha_{1}}
... {\left({\partial \over \partial x_{m}}\right)}^{\alpha_{m}}
{\left({\partial \over \partial \xi_{1}}\right)}^{\beta_{1}}
... {\left({\partial \over \partial \xi_{m}}\right)}^{\beta_{m}}
p^{0}(x,\xi) \right| \cr
& \leq c(x) \langle \xi \rangle^{d-|\beta|}.
&(36)\cr}
$$
We therefore come to appreciate the problematic aspect of symbols
of pseudo-differential operators.$^{12}$ The singularity at
$\xi=0$ can be dealt with either by modifying the leading symbol
for small $\xi$ to be a $C^{\infty}$ function (at the price of
loosing the homogeneity there), or by keeping the strict 
homogeneity and dealing with the singularity at $\xi=0.^{12}$

On the other hand, we are interested in a definition of strong
ellipticity of pseudo-differential boundary-value problems that
reduces to Eqs. (20)--(22) when both $P$ and the boundary 
operator reduce to the form considered in Ref. 6. For this
purpose, and bearing in mind the occurrence of singularities
in the leading symbols of $P$ and of the boundary operator,
we make the following requirements.$^{10}$

Let $(P+G)$ be a pseudo-differential operator subject to boundary
conditions described by the pseudo-differential boundary
operator $\cal B$ (the consideration of $(P+G)$ rather than only
$P$ is necessary to achieve self-adjointness, as is described in
detail in Refs. 12 and 13).
The pseudo-differential boundary-value problem
$((P+G),{\cal B})$ is strongly elliptic with respect to 
${\cal C}-{\bf R}_{+}$ if:
\vskip 0.3cm
\noindent
(I) The inequalities (30) and (34) hold;
\vskip 0.3cm
\noindent
(II) There exists a unique solution of the equation
$$
\left[\sigma_{L}\left((P+G); \left \{ {\hat x}^{k} \right \},r=0,
\left \{ \zeta_{j} \right \}, \omega \rightarrow 
-i{\partial \over \partial r} \right)-\lambda \right]
\varphi(r,{\hat x},\zeta;\lambda)=0,
\eqno (20')
$$
subject to the boundary conditions
$$
\sigma_{L}({\cal B})\left( \left \{ {\hat x}^{k} \right \},
\left \{ \zeta_{j} \right \} \right)\psi(\varphi)
=\psi'(\varphi)
\eqno (21')
$$
and to the asymptotic condition (22). It should be stressed that,
unlike the case of differential operators, Eq. (20') is not an
ordinary differential equation in general, because $(P+G)$ is
pseudo-differential.
\vskip 0.3cm
\noindent
(III) The strictly homogeneous symbols associated to $(P+G)$ and
$\cal B$ have limits for $|\zeta| \rightarrow 0$ in the respective
leading symbol norms, with the limiting symbol restricted to
the boundary which avoids the values $\lambda \not \in {\cal C}
-{\bf R}_{+}$ for all $\left \{ {\hat x} \right \}$.

Condition (III) requires a last effort for a proper understanding.
Given a pseudo-differential operator of order $d$ with leading
symbol $p^{0}(x,\xi)$, the associated strictly homogeneous symbol
is defined by$^{12}$
$$
p^{h}(x,\xi) \equiv |\xi|^{d} p^{0} \left(x,{\xi \over |\xi|}\right)
\; \; {\rm for} \; \; \xi \not = 0.
\eqno (37)
$$
This extends to a continuous function vanishing at $\xi=0$ when
$d>0$. In the presence of boundaries, the boundary-value problem
$((P+G),{\cal B})$ has a strictly homogeneous symbol on the
boundary equal to (some indices are omitted for simplicity)
$$
\pmatrix{p^{h}\left( \left \{ {\hat x} \right \},r=0,
\left \{ \zeta \right \},-i{\partial \over \partial r} \right)
+g^{h}\left( \left \{ {\hat x} \right \}, 
\left \{ \zeta \right \},-i{\partial \over \partial r}
\right)- \lambda \cr
b^{h} \left( \left \{ {\hat x} \right \}, 
\left \{ \zeta \right \}, -i{\partial \over \partial r}
\right) \cr},
$$
where $p^{h},g^{h}$ and $b^{h}$ are the strictly homogeneous 
symbols of $P,G$ and $\cal B$ respectively, obtained from the
corresponding leading symbols $p^{0},g^{0}$ and $b^{0}$ via
equations analogous to (37), after taking into account 
the split (19), and upon replacing $\omega$
by $-i{\partial \over \partial r}$. The limiting symbol restricted
to the boundary (also called limiting $\lambda$-dependent boundary
symbol operator) and mentioned in condition III reads therefore$^{12}$
$$ \eqalignno{
\; & a^{h} \left( \left \{ {\hat x} \right \}, r=0, \zeta=0,
-i{\partial \over \partial r} \right) \cr
&=\pmatrix{
p^{h} \left( \left \{ {\hat x} \right \}, r=0, \zeta=0,
-i{\partial \over \partial r} \right)
+g^{h} \left( \left \{ {\hat x} \right \}, \zeta=0,
-i{\partial \over \partial r} \right) -\lambda \cr
b^{h} \left( \left \{ {\hat x} \right \}, \zeta=0, 
-i{\partial \over \partial r} \right) \cr},
&(38)\cr}
$$
where the singularity at $\xi=0$ of the leading symbol in absence
of boundaries is replaced by the singularity at $\zeta=0$ of the
leading symbols of $P,G$ and $\cal B$ when a boundary occurs.  

Let us now see how the previous conditions on the leading symbol
of $(P+G)$ and on the graded leading symbol of the boundary operator
can be used. The equation (20') is solved by a function $\varphi$
depending on $r, \left \{ {\hat x}^{k} \right \}, 
\left \{ \zeta_{j} \right \}$ and, parametrically, on the eigenvalues
$\lambda$. For simplicity, we write
$\varphi=\varphi(r,{\hat x},\zeta;\lambda)$, omitting indices. Since 
the leading symbol is no longer a polynomial when $(P+G)$ is 
genuinely pseudo-differential, we cannot make any further specification
on $\varphi$ at this stage, apart from requiring that it should
reduce to (here $|\zeta|^{2} \equiv \zeta_{i}\zeta^{i}$)
$$
\chi({\hat x},\zeta)e^{-r\sqrt{|\zeta|^{2}-\lambda}}
$$
when $(P+G)$ reduces to a Laplacian. 

The equation (21') involves the graded leading symbol of $\cal B$
and restrictions to the boundary of the field and its covariant
derivative along the normal direction. Such a restriction is
obtained by setting to zero the geodesic distance $r$, and hence
we write, in general form (here we denote again by $\Lambda$ the
full matrix element ${\cal B}_{21}$ in the boundary operator (10)),
$$
\pmatrix{\Pi & 0 \cr \sigma_{L}(\Lambda) & I-\Pi \cr}
\pmatrix{\varphi(0,{\hat x},\zeta;\lambda) \cr
\varphi'(0,{\hat x},\zeta;\lambda) \cr}
=\pmatrix{\Pi \rho(0,{\hat x},\zeta;\lambda) \cr
(I-\Pi) \rho'(0,{\hat x},\zeta;\lambda) \cr},
\eqno (39)
$$
where $\rho$ differs from $\varphi$, because Eq. (21') is
written for $\psi(\varphi)$ and $\psi'(\varphi) \not =
\psi(\varphi)$. Now Eq. (39) leads to 
$$
\Pi \varphi(0,{\hat x},\zeta;\lambda)=\Pi 
\rho(0,{\hat x},\zeta;\lambda),
\eqno (40)
$$
$$
\sigma_{L}(\Lambda)\varphi(0,{\hat x},\zeta;\lambda)
+(I-\Pi)\varphi'(0,{\hat x},\zeta;\lambda)
=(I-\Pi)\rho'(0,{\hat x},\zeta;\lambda),
\eqno (41)
$$
and we require that, for $\varphi$ satisfying Eq. (20') and the
asymptotic decay (22), with $\lambda \in {\cal C}-{\bf R}_{+}$,
Eqs. (40) and (41) can be always solved with given values of
$\rho(0,{\hat x},\zeta;\lambda)$ and $\rho'(0,{\hat x},\zeta;\lambda)$,
whenever $(\zeta,\lambda) \not = (0,0)$. The idea is now to relate,
if possible, $\varphi'(0,{\hat x},\zeta;\lambda)$ to 
$\varphi(0,{\hat x},\zeta;\lambda)$ in such a way that Eq. (40) can
be used to simplify Eq. (41). For this purpose, we consider the
function $f$ such that
$$
{\varphi'(0,{\hat x},\zeta;\lambda)\over 
\varphi(0,{\hat x},\zeta;\lambda)}=
{\rho'(0,{\hat x},\zeta;\lambda)\over 
\rho(0,{\hat x},\zeta;\lambda)}=f({\hat x},\zeta;\lambda),
\eqno (42)
$$
$$
\Pi({\hat x})f({\hat x},\zeta;\lambda)=f({\hat x},\zeta;\lambda)
\Pi({\hat x}).
\eqno (43)
$$
If both (42) and (43) hold, Eq. (41) reduces indeed to
$$ \eqalignno{
\; & \sigma_{L}(\Lambda)\varphi(0,{\hat x},\zeta;\lambda)
+f({\hat x},\zeta;\lambda)\Bigr(\varphi(0,{\hat x},\zeta;\lambda)
-\rho(0,{\hat x},\zeta;\lambda)\Bigr) \cr
&=f({\hat x},\zeta;\lambda)\Pi \Bigr(\varphi(0,{\hat x},\zeta;\lambda)
-\rho(0,{\hat x},\zeta;\lambda) \Bigr),
&(44a)\cr}
$$
and hence, by virtue of (40), 
$$
\Bigr[\sigma_{L}(\Lambda)+f({\hat x},\zeta;\lambda)\Bigr]
\varphi(0,{\hat x},\zeta;\lambda)=\rho'(0,{\hat x},\zeta;\lambda).
\eqno (44b)
$$
Thus, the strong ellipticity condition with respect to 
${\cal C}-{\bf R}_{+}$ implies in this case the invertibility of
$\Bigr[\sigma_{L}(\Lambda)+f({\hat x},\zeta;\lambda)\Bigr]$, i.e.
$$
{\rm det} \Bigr[\sigma_{L}(\Lambda)+f({\hat x},\zeta;\lambda)
\Bigr] \not = 0 \; \; \; \; \forall \lambda \in 
{\cal C}-{\bf R}_{+}.
\eqno (45)
$$
Moreover, by virtue of the identity
$$
\Bigr[f({\hat x},\zeta;\lambda)+\sigma_{L}(\Lambda)\Bigr]
\Bigr[f({\hat x},\zeta;\lambda)-\sigma_{L}(\Lambda)\Bigr]
=\Bigr[f^{2}({\hat x},\zeta;\lambda)-\sigma_{L}^{2}(\Lambda)\Bigr],
\eqno (46)
$$
the condition (45) is equivalent to
$$
{\rm det}\Bigr[f^{2}({\hat x},\zeta;\lambda)-\sigma_{L}^{2}(\Lambda)
\Bigr] \not = 0 \; \; \; \; \forall \lambda \in 
{\cal C}-{\bf R}_{+}.
\eqno (47)
$$
Since $f({\hat x},\zeta;\lambda)$ is, in general, complex-valued, one
can always express it in the form
$$
f({\hat x},\zeta;\lambda)={\rm Re}f({\hat x},\zeta;\lambda)
+i{\rm Im}f({\hat x},\zeta;\lambda),
\eqno (48)
$$
so that (47) reads eventually
$$
{\rm det}\Bigr[{\rm Re}^{2}f({\hat x},\zeta;\lambda)
-{\rm Im}^{2}f({\hat x},\zeta;\lambda)-\sigma_{L}^{2}(\Lambda)
+2i {\rm Re}f({\hat x},\zeta;\lambda)
{\rm Im}f({\hat x},\zeta;\lambda)\Bigr] \not = 0.
\eqno (49)
$$
In particular, when
$$
{\rm Im}f({\hat x},\zeta;\lambda)=0,
\eqno (50)
$$
condition (49) reduces to
$$
{\rm det}\Bigr[{\rm Re}^{2}f({\hat x},\zeta;\lambda)
-\sigma_{L}^{2}(\Lambda)\Bigr] \not = 0.
\eqno (51)
$$
A {\it sufficient condition} for strong ellipticity with respect
to the cone ${\cal C}-{\bf R}_{+}$ is therefore the negative-definiteness
of $\sigma_{L}^{2}(\Lambda)$:
$$
\sigma_{L}^{2}(\Lambda) < 0,
\eqno (52)
$$
so that
$$
{\rm Re}^{2}f({\hat x},\zeta;\lambda)-\sigma_{L}^{2}(\Lambda)>0,
\eqno (53)
$$
and hence (51) is fulfilled.

In the derivation of the sufficient conditions (49) and (52), the
assumption (43) plays a crucial role. In general, however, $\Pi$
and $f$ have a non-vanishing commutator, and hence a
$C({\hat x},\zeta;\lambda)$ exists such that
$$
\Pi({\hat x})f({\hat x},\zeta;\lambda)
-f({\hat x},\zeta;\lambda)\Pi({\hat x})
=C({\hat x},\zeta;\lambda).
\eqno (54)
$$
The occurrence of $C$ is a peculiar feature of the fully
pseudo-differential framework. Equation (41) is then equivalent to
(now we write explicitly also the independent variables in the
leading symbol of $\Lambda$)
$$ \eqalignno{
\; & \Bigr[(\sigma_{L}(\Lambda)-C)({\hat x},\zeta;\lambda)
+f({\hat x},\zeta;\lambda)\Bigr]\varphi(0,{\hat x},\zeta;\lambda) \cr
&=\rho'(0,{\hat x},\zeta;\lambda)-C({\hat x},\zeta;\lambda)
\rho(0,{\hat x},\zeta;\lambda).
&(55)\cr}
$$
On defining
$$
\gamma({\hat x},\zeta;\lambda) \equiv \Bigr[\sigma_{L}(\Lambda)
-C \Bigr]({\hat x},\zeta;\lambda),
\eqno (56)
$$
we therefore obtain strong ellipticity conditions formally analogous
to (45) or (49) or (51), with $\sigma_{L}(\Lambda)$ replaced
by $\gamma({\hat x},\zeta;\lambda)$ therein, i.e.
$$
{\rm det}\Bigr[\gamma({\hat x},\zeta;\lambda)
+f({\hat x},\zeta;\lambda)\Bigr] \not = 0 \; \; \forall
\lambda \in {\cal C}-{\bf R}_{+},
\eqno (57)
$$
which is satisfied if
$$
{\rm det}\Bigr[{\rm Re}^{2}f({\hat x},\zeta;\lambda)
-{\rm Im}^{2}f({\hat x},\zeta;\lambda)-\gamma^{2}({\hat x},\zeta;\lambda)
+2i{\rm Re}f({\hat x},\zeta;\lambda){\rm Im}f({\hat x},\zeta;\lambda)
\Bigr] \not = 0.
\eqno (58)
$$
We have therefore provided a complete characterization of the
properties of the symbol of the boundary operator for which a set
of boundary conditions completely invariant under infinitesimal
diffeomorphisms are compatible with a strongly elliptic one-loop
quantum theory. Our analysis is detailed but general,
and hence has the merit (as far as we can see) of including all
pseudo-differential boundary operators for which the sufficient
conditions just derived can be imposed.

It would be now very interesting to prove that, by virtue of the
pseudo-differential nature of $\cal B$ in (10), the quantum state
of the universe in one-loop semiclassical theory can be made of
surface-state type.$^{14}$ This would describe a wave function of
the universe with exponential decay away from the boundary, which
might provide a novel description of quantum physics
at the Planck length. It therefore seems 
that by insisting on path-integral quantization, strong
ellipticity of the Euclidean theory and invariance principles,
new deep perspectives are in sight. These are in turn closer to what
we may hope to test, i.e. the one-loop semiclassical approximation
in quantum gravity. In the seventies, such calculations could
provide a guiding principle for selecting couplings of matter fields
to gravity in a unified field theory. Now they can lead instead to
a deeper understanding of the interplay 
between non-local formulations,$^{15-17}$
elliptic theory, gauge-invariant quantization$^{18}$ and a quantum 
theory of the very early universe.$^{10}$
\vskip 5cm
\leftline {\bf References}
\item {1.}
B. S. DeWitt, {\it Dynamical Theory of Groups and Fields}
(Gordon and Breach, New York, 1965).
\item {2.}
B. S. DeWitt, 
in {\it Relativity, Groups and Topology II}, eds. B. S. DeWitt
and R. Stora (North-Holland, Amsterdam, 1984). 
\item {3.}
R. P. Feynman, {\it Rev. Mod. Phys.} {\bf 20}, 367 (1948).
\item {4.}
C. W. Misner, {\it Rev. Mod. Phys.} {\bf 29}, 497 (1957).
\item {5.}
S. W. Hawking, 
in {\it General Relativity, an Einstein Centenary Survey}, eds. 
S. W. Hawking and W. Israel (Cambridge University Press, 
Cambridge, 1979). 
\item {6.}
I. G. Avramidi and G. Esposito, {\it Commun. Math. Phys.} 
{\bf 200}, 495 (1999).
\item {7.}
G. Grubb, {\it Ann. Scuola Normale Superiore Pisa} {\bf 1},
1 (1974).
\item {8.}
P. B. Gilkey and L. Smith, {\it J. Diff. Geom.} {\bf 18}, 393 (1983).
\item {9.}
G. Esposito, {\it Class. Quantum Grav.} {\bf 16}, 3999 (1999).
\item {10.}
G. Esposito, `Boundary Operators in Quantum Field Theory'
(HEP-TH 0001086).
\item {11.}
P. B. Gilkey, {\it Invariance Theory, the Heat Equation and the
Atiyah--Singer Index theorem} (Chemical Rubber Company,
Boca Raton, 1995).
\item {12.}
G. Grubb, {\it Functional Calculus of Pseudodifferential
Boundary Problems} (Birkh\"{a}user, Boston, 1996).
\item {13.}
G. Esposito, {\it Class. Quantum Grav.} {\bf 16}, 1113 (1999).
\item {14.}
M. Schr\"{o}der, {\it Rep. Math. Phys.} {\bf 27}, 259 (1989).
\item {15.}
J. W. Moffat, {\it Phys. Rev.} {\bf D41}, 1177 (1990).
\item {16.}
D. Evens, J. W. Moffat, G. Kleppe and R. P. Woodard,
{\it Phys. Rev.} {\bf D43}, 499 (1991).
\item {17.}
V. N. Marachevsky and D. V. Vassilevich, {\it Class. Quantum Grav.}
{\bf 13}, 645 (1996).
\item {18.}
G. Esposito and C. Stornaiolo, {\it Int. J. Mod. Phys.} {\bf A15},
449 (2000).

\bye